\documentclass{PoS}
\usepackage{graphicx}
\usepackage{amsmath}

\title{QCD phase diagram: an overview}

\ShortTitle{QCD phase diagram: an overview}

\author{\speaker{M. Stephanov}\\
        Physics Department, University of Illinois, 845 W. Taylor St.,
        Chicago, IL 60607-7059, USA\\
        E-mail: \email{misha@uic.edu}}

\abstract{
The aim of this review is to summarize
 the contemporary understanding of the
QCD phase diagram as a function of temperature $T$ and baryo-chemical
potential $\mu_B$. The focus is on recent theoretical developments 
due to lattice simulations of the phase diagram. 
}

\FullConference{XXIV International Symposium on Lattice Field Theory\\
		 July 23-28 2006\\
		 Tucson Arizona, US}

\newcommand\Nf{N_{\rm f}}
\DeclareMathOperator{\const}{const}

\begin{document}

\section{Introduction}
\label{sec:introduction}

Quantum Chromodynamics is a remarkable theory. It is a convincing
practical example of the triumph of the quantum field theory.
Asymptotic freedom allows QCD to be consistent down to
arbitrary short distance scale, enabling us to define the theory
completely in terms of the fundamental microscopic degrees of freedom
-- quarks and gluons. This fundamental definition is very simple, yet
the theory describes a wide range of phenomena -- from the mass
spectrum of hadrons to deep-inelastic processes. 
As such, QCD should
also possess well defined thermodynamic properties.
The knowledge of QCD thermodynamics is essential for the understanding
of such natural phenomena as compact stars
and laboratory experiments involving relativistic heavy-ion collisions.

Full analytical treatment of QCD is very difficult because,
neglecting quark masses, this theory has no numerically small
fundamental parameters. The only independent intrinsic scale in this
theory is the dynamically generated confinement scale $\Lambda_{\rm
  QCD}\sim 1 {\rm fm}^{-1}$.  
In certain limits, in particular, for
large values of the external thermodynamic parameters temperature $T$
and/or baryo-chemical potential $\mu_B$, when thermodynamics is
dominated by short-distance QCD dynamics, the theory can be studied
analytically, due to the asymptotic freedom. But, as seen below, the
most interesting experimental region of parameters $T$ and $\mu_B$ is
that of order $\Lambda_{\rm QCD}$.

The above makes first principle lattice approaches, which do not rely
on small parameter expansions, an invaluable and the most powerful tool
in studying QCD thermodynamics.  In addition, the domain where all
relevant time/distance (or energy/momentum) scales are similar is
especially suited for a lattice study (accommodating a wide scale
window would require a correspondingly large lattice).

The full potential of lattice methods is close to being
realized as far as the study of QCD at $\mu_B=0$ is concerned.  The
main practical problems in this regime -- accommodating sufficiently
light quarks and approaching the continuum limit -- are being
methodically and successfully addressed through the use of
improved lattice discretization schemes, as well as advances in
algorithm and hardware technology.

The status of thermodynamics of QCD at {\em non-zero} 
$\mu_B$ is different. The main impediment to lattice simulations is
the notorious sign problem, discussed in Section~\ref{sec:import-sampl-sign}.
No method devised so far is known, or expected, to converge
to the correct physical result as the infinite volume limit is approached at
{\em fixed} $\mu_B\ne0$. However, since the most interesting structure
of the QCD phase diagram (phase transitions and critical points)
lie at nonzero $\mu_B$, any progress in this direction is especially
valuable. Existing lattice methods generically rely on
clever extrapolations from \mbox{$\mu_B=0$}. These techniques yield interesting
results in the regime of small, but already experimentally relevant
$\mu_B$.

A contemporary view of the QCD phase diagram is shown in
Fig.~\ref{fig:pdqcd}. It is a compilation of a body of results from
model calculations, empirical nuclear physics, as well as first
principle lattice QCD calculations and perturbative calculations in
asymptotic regimes.

Several reviews in these proceedings, in addition to original
contributions, are devoted to recent progress in lattice studies of
QCD thermodynamics. Ref.~\cite{heller} reviews thermodynamics of
QCD at  $\mu_B=0$. Ref.~\cite{schmidt} discusses lattice results
at small $\mu_B$. In addition, Ref.\cite{alford} describes resent
progress in uncovering phase structure of QCD at large $\mu_B$,
relevant to the physics of compact stars, and outlines targets of
opportunity for potential future lattice studies in this domain.

This report provides an overview of the structure of the
QCD phase diagram based on available theoretical (lattice
and model calculations) and
phenomenological input.
Some of the recent lattice results reported separately 
in this volume are also briefly discussed in Section~\ref{sec:lattice-results}.

\section{The phase diagram}
\label{sec:basic}

Thermodynamic properties of a system are most readily expressed in
terms of a phase diagram in the space of thermodynamic parameters
-- in the case of QCD -- as a $T \mu_B$ phase diagram. Each point on the
diagram corresponds to a stable thermodynamic state, characterized
by various thermodynamic functions, such as, e.g., pressure, 
baryon density, etc (as well as kinetic coefficients, e.g., diffusion or
viscosity coefficients, or other properties of various correlation
functions).

Static thermodynamic quantities can be derived from the
partition function -- a Gibbs sum over eigenstates of QCD Hamiltonian,
which can be alternatively expressed as a path integral in
Euclidean space:
\begin{equation}
  \label{eq:Z-Gibbs}
  Z(T,\mu_B)=\sum_\alpha \exp\left\{-\frac{E_\alpha-\mu_B B_\alpha}T\right\}
=
\int{\cal D}(A,q,q^\dag) \exp\left\{-S_E\right\}
\end{equation}
where $\alpha$ labels states with energy $E_\alpha$ and baryon number
$B_\alpha$.
The path integral is over color gauge (gluon) fields $A_\mu$
periodic in Euclidean time with period $1/T$, and quarks fields,
antiperiodic with the same period.
The Euclidean action given by
\begin{equation}
  \label{eq:S-E}
  S_E=S_{\rm YM} + \sum_{q=u,d,s} \int d^4x \ q^\dag\, \mathbb D\, q 
\end{equation}
where $S_{\rm YM}[A]$ is the SU(3) Yang-Mills action and,
in the chiral Weyl basis, the Dirac spinors and matrix are
\begin{equation}
  \label{eq:Dirac-matrix}
  q=\left(
    \begin{array}{c}
      q_L\\q_R
    \end{array}
\right)
\quad{\rm and}\quad
\mathbb D = \left(
    \begin{array}{cc}
      \sigma\cdot D & m_q\\m_q & \sigma^\dag\cdot D\\
    \end{array}
\right)
- \mu_q
\end{equation}
where $\sigma_\mu=(1,i\boldsymbol{\sigma})$,
 $D_\mu=\partial_\mu-iA_\mu$, and $\mu_q=\mu_B/3$.

\subsection{Chiral symmetry argument}
\label{sec:chir-symm-argum}

In the chiral limit -- the idealized limit when 2 lightest quarks, $u$ and $d$,
are taken to be massless, the Lagrangian of QCD acquires 
chiral symmetry SU(2)$_L\times$SU(2)$_R$,
corresponding to SU(2) flavor rotations of $(u_L,d_L)$ and $(u_R,d_R)$
doublets independently. The ground state of QCD breaks the chiral symmetry
spontaneously locking SU(2)$_L$ and SU(2)$_R$ rotations into a single 
vector-like SU(2)$_V$ (isospin) symmetry and
generating 3 massless Goldstone pseudoscalar bosons -- the pions.  
The breaking of the chiral symmetry is a non-perturbative phenomenon.  

At sufficiently high temperature $T\gg\Lambda_{\rm QCD}$, due to the
asymptotic freedom of QCD, perturbation theory around the
approximation of the gas of free quarks and gluons (quark-gluon plasma
-- QGP) should become applicable. In this regime chiral symmetry is
not broken. Thus we must expect a transition from a broken chiral
symmetry vacuum state to a chirally symmetric equilibrium state at
some temperature $T_c\sim\Lambda_{\rm QCD}$. The transition is akin to
the Curie point in a ferromagnet -- where the rotational O(3) symmetry
is restored by thermal fluctuations (chiral O(4)=SU(2)$\times$SU(2)
symmetry in QCD).  Thermodynamic functions of QCD must be singular at
the transition point -- as always when the transition separates
thermodynamic states of different global symmetry.

\begin{figure}
  \centering
\includegraphics[width=.8\textwidth]{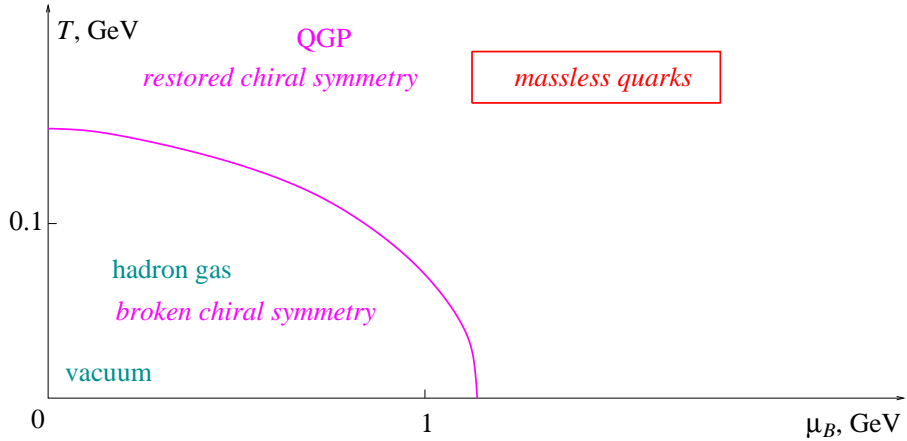}
\caption[]{Phase diagram of QCD with massless quarks
dictated by the chiral symmetry argument. The order of the
transition (solid magenta line) is not determined by this simplest argument.}
\label{fig:pdqcd-0}
\end{figure}

Thus, the region of broken chiral symmetry on the
$T \mu_B$ phase diagram must be separated from the region of the
restored symmetry by a closed boundary as shown in 
Fig.~\ref{fig:pdqcd-0}.

\subsection{Pisarski-Wilczek argument}
\label{sec:pw-argument}

The chiral symmetry argument alone is not sufficient to determine the
{\em order} of the temperature driven chiral symmetry restoration
transition. A more elaborate argument, based on universality, advanced
by Pisarski and Wilczek \cite{PisarskiWilczek} asserts that the
transition {\em cannot} be of second order for {\em three massless}
quarks.

In a simplified form, the logic of Ref.\cite{PisarskiWilczek} is
as follows. Let us assume that the transition is of the second order.
Then the critical behavior of the system (long-distance behavior of
correlation functions, singular contributions to thermodynamic
functions, etc.) is determined by the long-wavelength modes
which, in the case of the second order transition in a theory with
$\Nf$ light quarks, are the $\Nf^2-1$ pions of the spontaneously broken
SU$(\Nf)_A$ axial flavor symmetry plus the critical mode -- the
magnitude of the chiral condensate $\sigma\sim\bar qq$.

Universality implies that the critical behavior is the same as in any local
theory in 3 dimensions with the same global symmetry breaking pattern
and the same set of critical modes. In our case, a representative
example of the universality class is an SU($\Nf$)$\times$ SU($\Nf$)
sigma model of an $\Nf\times \Nf$ matrix-valued field $\Phi$. It turns
out, that for $\Nf=3$, the model cannot be critical: there is a
relevant operator cubic in the order parameter field, $\det\Phi$, which
always destabilizes the symmetric minimum of the effective potential
for $\Phi$ via a first order transition, before the curvature of the
minimum vanishes (i.e., before criticality is reached). 
Hence, QCD with $\Nf=3$ massless quark flavors must
undergo a {\em first} order chiral restoration transition.

\subsection{$\Nf=2$ chiral limit and tricritical point}

\begin{figure}
  \centering
\includegraphics[width=.8\textwidth]{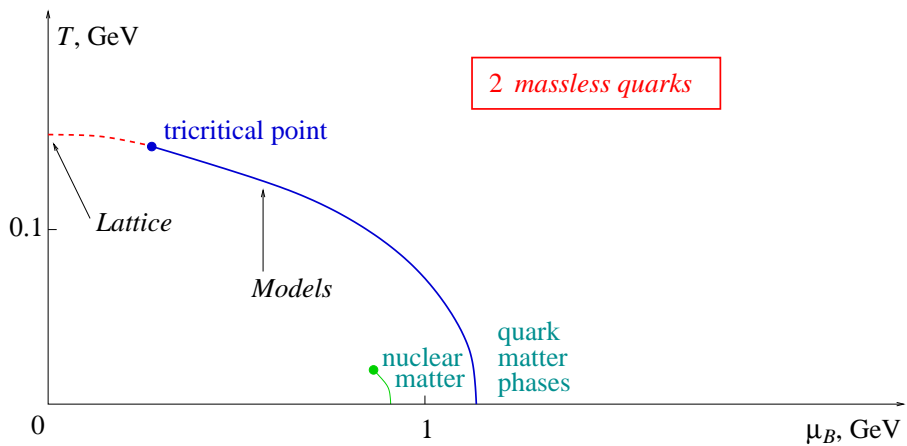}
\caption[]{The order of the chiral symmetry restoration
transition for 2 massless and one massive quarks. The dashed line
(red) is the second order transition, the solid line (blue) is the
first order transition. In the low $T$ region:
chiral symmetry is broken in nuclear matter. Details of the
phase structure at high $\mu_B$ are omitted.}
\label{fig:pdqcd-1}
\end{figure}

For {\em two massless} quarks the transition can be either second or first
order.  As lattice and model calculations show, both possibilities are
realized depending on the value of the strange quark mass $m_s$ and/or
the baryo-chemical potential $\mu_B$.

The point on the chiral phase transition line where the transition
changes order is called tricritical point, see Fig.~\ref{fig:pdqcd-1}.
The location of this point is one of the unknowns of the QCD phase
diagram with 2 massless quarks.  In fact, even the order of the
transition at $\mu_B=0$, which many older and recent studies
suggest
is of the second order (as shown in Fig.~\ref{fig:pdqcd-1})
is still being questioned (see review by Heller in this volume
\cite{heller}). 

Neither can it be claimed reliably (model or assumption independently)
that the transition, if it begins as a 2nd order at $\mu_B=0$, changes
to first order. However, numerous model calculations show this is the
case (Section \ref{sec:models}).  Lattice calculations also support
such a picture. Recent advances in the understanding of QCD at low $T$
and large $\mu_B$, reviewed in \cite{alford}, also point at a first
order transition (at low-$T$, high-$\mu_B$) from nuclear matter to
color-superconducting quark matter phase.  Fig.~\ref{fig:pdqcd-1}
reflects this consensus.

At low temperature, nuclear matter (which is expected to be still
bound in the chiral limit) should be placed on the broken symmetry
side of the chiral transition line as shown in Fig.~\ref{fig:pdqcd-1}.

\subsection{Physical quark masses and crossover}

\begin{figure}
  \centering
\includegraphics[width=.8\textwidth]{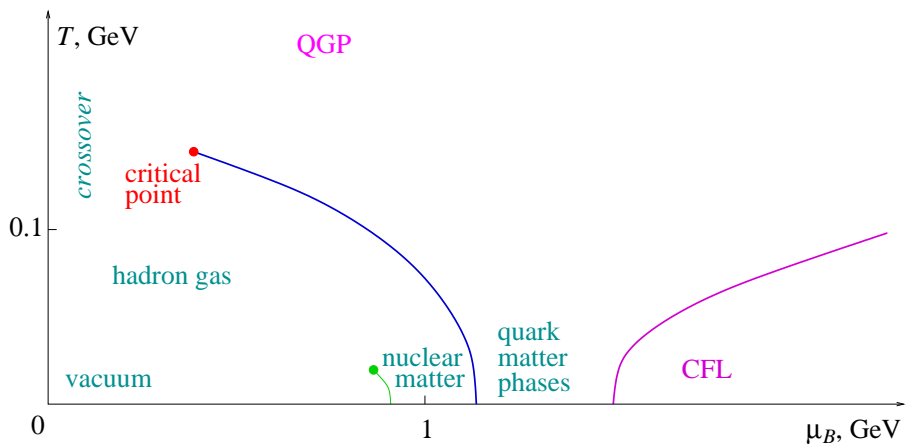}
\caption[]{The contemporary view of the QCD phase diagram -- a
  semiquantitative sketch.
}
\label{fig:pdqcd}
\end{figure}

When the up and down quark masses are set to their observed finite 
values, the diagram assumes the shape sketched in Fig.~\ref{fig:pdqcd}. The
second order transition line (where there was one)
is replaced by a crossover -- the
criticality needed for the second order transition in
Fig.~\ref{fig:pdqcd-1} requires tuning chiral symmetry breaking
parameters (quark masses) to zero.  In the absence of the exact chiral
symmetry (broken by quark masses) the transition from low- to 
high-temperature phases of QCD need not proceed through a singularity.
Lattice simulations do indeed show that the transition is a crossover
 for $\mu_B=0$ (most recently and decisively Ref.\cite{szabo}, see also
Ref.\cite{heller} for a review).%
\footnote{This fact is technically easier to establish than the order of the 
transition in the chiral limit
  -- taking the chiral limit is an added difficulty.}

This transitional crossover region is notoriously
difficult to describe or model analytically -- description in terms of the
hadronic degrees of freedom (resonance gas) breaks down as one
approaches crossover temperature (often called $T_c$), and the dual
description in terms of weakly interacting 
quarks and gluons does not become valid until much higher temperatures.
Recent terminology for the QCD state
near the crossover ($T\sim(1-2)T_c$) is
strongly coupled quark-gluon plasma (sQGP). 

Transport properties of
sQGP have attracted considerable attention. For example, generally, 
the shear viscosity $\eta$ is a decreasing function of the coupling
strength. The dimensionless ratio of $\eta/\hbar$ to the entropy density $s$
 tends to infinity asymptotically far on either side of
the crossover -- in dilute hadron gas ($T\to0$) 
and in asymptotically free QGP ($T\to\infty$).
Near the crossover $\eta/s$ should thus be expected to reach a minimum
\cite{Csernai:2006zz}.
The viscosity can be indirectly determined in heavy ion collisions
by comparing hydrodynamic calculations to experimental data.
Such comparison~\cite{Teaney:2003kp} 
indeed indicates that the viscosity (per entropy
density) of this ``crossover liquid'' is relatively small, and
plausibly is saturating
the lower bound conjectured in \cite{Kovtun:2004de}.

\subsection{Physical quark masses and the critical point}

The first order transition line is now ending at a point known as the
QCD critical point or end point.\footnote{The QCD critical point is
  sometimes also referred to as {\em chiral} critical point which sets
  it apart from another known (nuclear) critical point, the end-point
  of the transition separating nuclear liquid and gas phases (see
  Fig.~\ref{fig:pdqcd}).  This point occurs at much lower temperatures
  ${\cal O}(10 MeV)$ set by the scale of the nuclear binding
  energies.}
The end point of a first order line is a critical point of the second
order.  This is by far the most common critical phenomenon in
condensed matter physics.  Most liquids possess
such a singularity, including water. The line
which we know as the water boiling transition ends at pressure $p=218$ atm
and $T=374^\circ$C.  Along this line the two coexisting phases (water and
vapor) become less and less distinct as one approaches the end point
(the density of water decreases and of vapor increases),
resulting in a single phase at this point and beyond.

In QCD the two coexisting phases are hadron gas (lower $T$), and
quark-gluon plasma (higher $T$). What distinguishes the two phases? As
in the case of water and vapor, the distinction is only quantitative,
and more obviously so as we approach the critical point. Since the
chiral symmetry is explicitly broken by quark masses, the two phases
cannot be distinguished by realizations (broken vs restored)
of any global symmetry.\footnote{Deconfinement, although a useful
  concept to discuss the transition from hadron to quark-gluon plasma,
  strictly speaking, does not provide a distinction between the phases.
  With quarks, even in vacuum ($T=0$) the confining potential cannot
  rise infinitely -- a quark-antiquark pair inserted into the color
  flux tube breaks it.  The energy required to separate two test color
  charges from each other is finite if there are dynamical quarks.}

It is worth pointing out that beside the critical point, the phase
diagram of QCD in Fig.~\ref{fig:pdqcd} has other similarities with the
phase diagram of water.  A number of ordered quark matter phases must
exist in the low-$T$ high-$\mu_B$ region, which are akin to many (more
than 10) confirmed phases of ice.  For asymptotically large $\mu_B$,
QCD with 3 quark flavors must be in color-flavor locked (CFL)
state~\cite{CFL,alford}.

\section{Locating the critical point: theory}

The critical point is a well-defined singularity on the phase diagram,
and it appears as an attractive theoretical, as well as experimental,
target to shoot at.
Theoretically, finding the coordinates $(T,\mu_B)$ of the
critical point is a straightforwardly defined task. We need to calculate
the partition function of QCD given by Eq.~(\ref{eq:Z-Gibbs})
 and find the singularity corresponding
to the end of the first order transition line. But it is easier said
than done.

Of course, calculating such an infinitely
dimensional integral analytically is beyond present reach.
Numerical lattice Monte Carlo simulations is an obvious
tool to choose for this task. At zero $\mu_B$
Monte Carlo method allows us to determine the equation of state of QCD
as a function of $T$ (and show that the transition is a crossover). 
However, at {\em finite} $\mu_B$ the Nature guards its secrets better.

\subsection{Importance sampling and the sign problem}
\label{sec:import-sampl-sign}

The notorious sign problem has been known to lattice Monte Carlo experts since
the early days of this field. Calculating the partition function using
Monte Carlo method hinges on the fact that the exponent of the
Euclidean action $S_E$ is a positive-definite function of its 
variables (values of the fields on the lattice). This allows
one to limit calculation to a relatively small set of field configurations
randomly picked with probability proportional to the value of $\exp(-S_E)$.
The number of such configurations needed to achieve reasonable
accuracy is vastly smaller than the total number of possible
configurations. The latter is exponentially large in the size $V$ 
of the system, 
or, the number of the degrees of freedom: $\exp(\const \cdot V)$.
The method, also known as importance sampling, utilizes the fact
that the vast majority of these configurations contribute a
tiny fraction because of the exponential suppression by $\exp(-S_E)$.
Only configurations with sizable $\exp(-S_E)$ are important.

In QCD with $\mu_B\neq0$ the Monte Carlo action $S_{\rm MC}$ (playing
the role of $S_E$) is complex.  With $S_{\rm MC}$ complex, how does
one pick important configurations? A number of ways to circumvent the
problem have been tried. For example, using the modulus of
$\exp(-S_{\rm MC})$ as a measure of importance, or the value of
$\exp(-S_{\rm MC})$ at {\em zero} $\mu_B$, when it is still positive.
Unfortunately, none of the methods can be expected to converge to
correct result with the increasing lattice volume, unless this limit
is not accompanied by an {\rm exponential} $\exp(\const \cdot V)$
increase of the number of configurations, rendering Monte Carlo
technique useless.

\subsection{The overlap problem}

To demonstrate the problem, consider the most straightforward
attempt to circumvent it -- reweighting.\footnote{For QCD at finite
  $\mu_B$ this method is known as the
  ``Glasgow method'' (reviewed in Ref. \cite{Barbour-review}).}
We cannot obtain
a correctly weighted sample of important configurations directly
at $\mu_B\ne0$. But we still can at $\mu_B=0$. So, we take the
$\mu_B=0$ sample
and offset incorrect probability by multiplying the contribution
of each configuration
by a factor $\exp(+S_E|_{\mu_B=0}-S_E)$. This is
exact in the limit when the sample contains {\em all} possible
configurations. 

The problem is in the size of the sample needed
for a Monte Carlo computation as $V\to\infty$. The method uses the
fact that at {\em finite} volume $V$, even at $\mu_B=0$, the
configurations important for ${\mu_B\ne0}$ pop up, but with a very
small probability.  This probability is exponentially small as volume
$V\to\infty$: $\exp(-{\rm const}\cdot V)$. When we calculate the
partition function the reweighting factor is correcting for that, and
is therefore exponentially large (for the complex $S_{\rm MC}$, both
he magnitude and the complex phase are).  Fluctuations, or statistical
noise, in the exponentially tiny number of the rare important
configurations completely washes out the significance of the result.

In layman's terms, imagine that we want to study ice, but can only run
experiments at normal room temperature and pressure. Using the
reweighting method is analogous to trying to glimpse the information
by waiting for rare configurations when all the water molecules
accidentally gather in one corner of the lab, forming a chunk of
ice. The amount of time that this experiment would require is
exponentially large as~$V\to\infty$.

\subsection{Complex determinant}
\label{sec:complex-determinant}

Why is the Monte Carlo action in QCD complex and what can be done
about it?
To see, integrate over the quark fields in (\ref{eq:Z-Gibbs})
explicitly and obtain
\begin{equation}
  \label{eq:Z-det}
  Z = \int{\cal D}A\ e^{-S_{\rm YM}}\prod_q \det\mathbb{D}
\equiv \int{\cal D}A\ e^{-S_{\rm MC}}
\end{equation}
where (as in Eq.~(\ref{eq:Dirac-matrix}) and using the property $D^\dag=-D$):
\begin{equation}
  \label{eq:D}
  \mathbb{D} = \left(
    \begin{array}{cc}
      \sigma\cdot D -\mu_q & m_q\\m_q & -(\sigma\cdot D)^\dag-\mu_q\\
    \end{array}
\right)
\end{equation}
For $\mu_q=0$ each quark determinant in Eq.~(\ref{eq:Z-det})
 is manifestly positive:
 \begin{equation}
   \label{eq:det-positive}
   \det\mathbb
D = \det[(\sigma\cdot D)(\sigma\cdot D)^\dag + m_q^2]>0.
 \end{equation}
The positivity (and even reality) is
lost if $\mu_q\ne0$. This is the sign problem.

However, the following still holds
 $(\det\mathbb
D(\mu_q))^*=\det\mathbb D(-\mu_q^*)$. 
This opens two possibilities for the measure in the Euclidean path
integral (\ref{eq:Z-det}) to remain positive for  $\mu_q\ne0$: 
(a) if  $\mu_q$ is imaginary; or (b) if there are two degenerate quarks,
e.g., $m_u=m_d$ and  $\mu_u=-\mu_d$, which is what happens with the chemical
potential $\mu_I$ of isospin $I_3$, or in phase-quenched QCD. 
Both alternatives are being exploited to glimpse into
the regime $\mu_B\neq0$, yet unaccessible to direct Monte Carlo.
In particular, the recent results from the simulations
at finite $\mu_I$ are reported in  Ref.\cite{sinclair}. Simulations
at imaginary $\mu_B$ are discussed further 
in Section~\ref{sec:imaginary-mu_b-nf3}.

\subsection{Predictions from models}
\label{sec:models}

\begin{figure}
  \centering
\includegraphics[width=.8\textwidth]{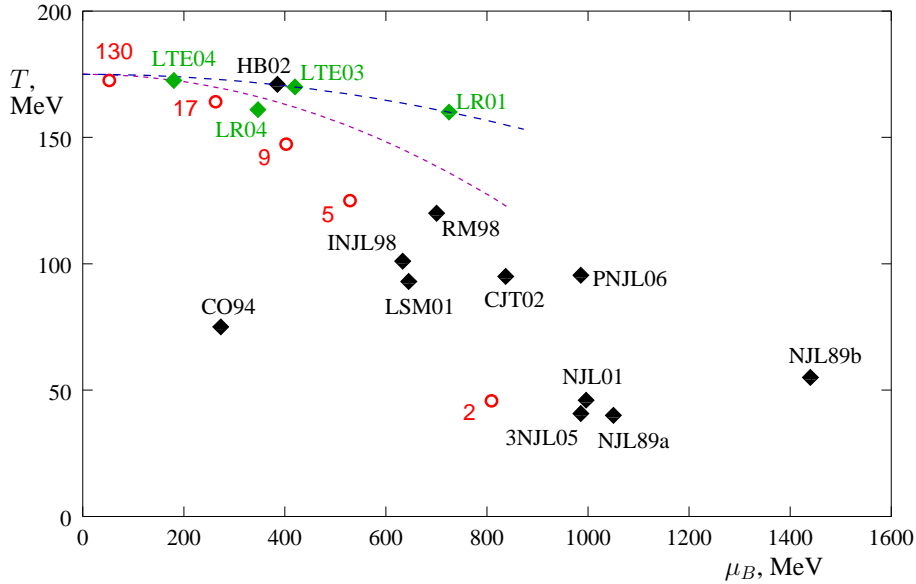}
\caption[]{Comparison of predictions for the location of the 
QCD critical point on the phase diagram. Black points are model predictions: 
NJLa89, NJLb89 --
\cite{Asakawa}, CO94 -- \cite{Barducci:1989,Barducci:1993}, INJL98
-- \cite{BergesRajagopal}, RM98 -- \cite{Halasz}, LSM01, NJL01 -- 
\cite{Scavenius}, HB02 -- \cite{Antoniou}, CJT02 -- \cite{HattaIkeda},
3NJL05 -- \cite{Barducci:2005ut}, PNJL06 -- \cite{Roessner:2006xn}.
Green points are lattice predictions: LR01, LR04 -- \cite{FK}, LTE03
-- \cite{Ejiri}, LTE04 -- \cite{Gavai}. The two dashed lines are
parabolas with slopes corresponding to lattice predictions of the
slope $dT/d\mu_B^2$ of the transition line at $\mu_B=0$
\cite{Ejiri,Philipsen}.
The red circles are locations of the freezeout points for heavy ion
collisions at corresponding center of mass
energies per nucleon
(indicated by labels in GeV) -- Section~\ref{sec:scanning-qcd-phase}.
}
\label{fig:tmudat} 
\end{figure}

In the absence of a controllable (i.e., systematically improvable and
converging in the \mbox{$V\to\infty$} limit) method to simulate QCD at
nonzero $\mu_B$, one turns to model calculations. Many such
calculations have been done
\cite{Asakawa,Barducci:1989,Barducci:1993,BergesRajagopal,Halasz,Scavenius,Antoniou,HattaIkeda,Barducci:2005ut,Roessner:2006xn}.
Figure \ref{fig:tmudat}
summarizes the results. One can see that the predictions vary
wildly. An interesting point to keep in mind is 
that each of these models is tuned to
reproduce vacuum, $T=\mu_B=0$, phenomenology.
Nevertheless, extrapolation to nonzero $\mu_B$ 
is not constrained significantly by this. In a loose sense,
most lattice methods (see next Section) can be also viewed as extrapolations
from $\mu_B=0$, albeit with reliable input from {\em finite} $T$.

\section{Lattice results on the critical point}
\label{sec:lattice-results}

This section is devoted to brief (and necessarily incomplete)
descriptions of currently developed lattice methods for reaching out
into the $T\mu_B$ plane. The comments below are selective and
are meant to complement the
original contributions in this volume.  For
a more comprehensive description of these methods, as well as other
methods not discussed here, the reader may consult
the most up-to-date review of Schmidt in these proceedings
\cite{schmidt} as well as an 
earlier review by Philipsen \cite{Philipsen-review}, both of which also
contain further references to original papers.

\subsection{Reweighting}
\label{sec:reweighting}

The first lattice prediction for the location of the critical point
was reported by Fodor and Katz in Ref.~\cite{FK}. 
The assumption is that, although
the problem becomes exponentially difficult as $V\to\infty$,
in practice, one can get a sensible approximation at finite $V$. 
In addition, simulations at finite $T$
might suffer lesser overlap problem because of large thermal 
fluctuations \cite{AKW}.
One can hope that if the critical point is at a small value of
$\mu_B$, the volume $V$ may not need to be too large
to achieve a reasonable accuracy. In particular, numerical estimates show
\cite{overlap} that the maximal value of $\mu_B$ which
one can reach within the same accuracy shrinks only as a power of
$1/V$.

The results of Ref.~\cite{FK} are the most definitive and well-known,
but they also attract the strongest of criticisms.  The method of the
Ref.~\cite{FK} is based on computing the position of the zero of the
partition function in the complex temperature plane and observing when
(for which $\mu_B$) this zero crosses (and with its complex conjugate
-- pinches) the real axis. This determines the $T$ and $\mu_B$
coordinates of the critical point. However, as Ejiri points out in
Ref.~\cite{Ejiri:2005ts}, once the fluctuations of the phase,
$\arg\det{\mathbb D}$, of the Dirac determinant are large, they cause
fake zeros to appear. It is therefore alarming that, as Splittorff
argues \cite{splittorff}, both points found in Ref.~\cite{FK}
(different $m_q$ and~$V$) happen to lie on the critical line of the
phase quenched QCD ($|\det{\mathbb D}|$ instead of $\det{\mathbb D}$)
-- which is the line where fluctuations of $\arg\det{\mathbb D}$ do
become large.  In a related observation, Golterman {\it et
  al}~\cite{Golterman:2006rw,svetitsky} argue that the procedure of
taking the fourth root \cite{sharpe} of the staggered fermions causes
problems in a finite $\mu_B$ calculation such as in Ref.\cite{FK}.

\subsection{Imaginary $\mu_B$ and  $\Nf=3$}
\label{sec:imaginary-mu_b-nf3}

By the universality argument of Section~\ref{sec:pw-argument},
the finite temperature transition is 1st order for $m_u=m_d=m_s=0$.
By continuity, it must remain 1st order in a finite domain
of the $m_s m_{ud}$ plane (taking $m_u=m_d=m_{ud}$)
surrounding the origin --
the plot of this domain is known as Columbia plot \cite{columbia,heller}.
For physical quark masses and $\mu_B=0$ the temperature driven
transition is a crossover,
which means that the physical point is outside the
1st order domain in the $m_s m_{ud}$ plot. Reducing quark masses
should pull the point into the 1st order domain.

\begin{figure}
  \centering
\includegraphics[width=0.8\textwidth]{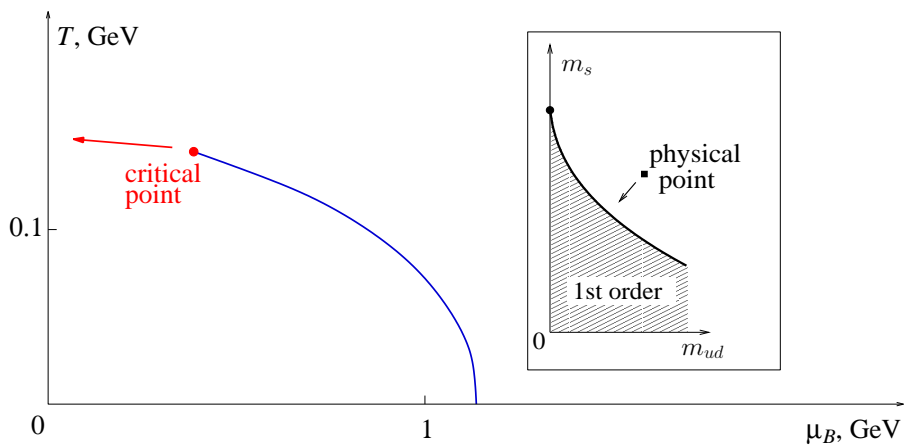}
  \caption{Expected direction of motion of the critical point (left) as the
    quark masses are decreased as shown in the Columbia plot in the inset.
}
  \label{fig:std-scenario}
\end{figure}

What happens on the $T\mu_B$ phase diagram as the point $m_sm_{ud}$
is pulled towards and into the first order domain?  The most straightforward
expectation is that the first order line begins earlier, at lower
$\mu_B$, i.e., the critical point is pulled towards the $\mu_B=0$ axis,
as shown in Fig.~\ref{fig:std-scenario},
until it disappears off the phase diagram altogether, and the whole
transition line is of the 1st order.\footnote{One can see how
  this scenario is realized on a 3-flavor NJL model 
  in Ref.\cite{Barducci:2005ut}.
}
Furthermore, lattice calculations at
$\mu_B=0$ show that real QCD is very near the 1st order domain
boundary. That suggests the critical point is not too far off the $T$
axis in
the $T\mu_B$ plane.

What happens to the critical point when $(m_s,m_{ud})$ is in the 1st
order domain? It is still a singularity of the partition function
as a function of $\mu_B$, but it moves out into the complex $\mu_B$
plane. More precisely, it moves to {\em imaginary\/} $\mu_B$ axis.
This remarkable fact allows one to observe the (complex descendant of)
the critical point in a direct Monte Carlo
simulation -- since there is no sign problem for imaginary $\mu_B$
(Section~\ref{sec:complex-determinant}). This observation is at the core of 
the method developed
by de Forcrand and Philipsen 
\cite{Philipsen,deforcrand}.

The success of the method crucially depends on the {\em analyticity}
of the coordinate $\mu_B^2$ of the critical point as a function of
quark mass, e.g., $m_s$ around the point where $\mu_B^2=0$.
The validity of this can be argued as follows.
In the $(T, \mu_B, m_s)$ space the criticality is achieved
(correlation length goes to infinity) when 2 conditions are
satisfied: $t(T, \mu_B^2, m_s)=h(T,\mu_B^2, m_s)=0$, i.e.,
 there are two relevant operators in the universality
class of the critical point and their coefficients, $t$ and $h$, must be
tuned to zero.
The coefficients of these
operators are analytic functions of the parameters.\footnote{The
non-analyticity characteristic of the critical behavior, arises due
to non-analytic dependence of the correlation length, $\xi$, on the
values of the relevant parameters $t$ and $h$.}  Furthermore,
the analyticity
in $\mu_B^2$ (not just in $\mu_B$, which otherwise could cause a
branching point at $\mu_B^2=0$) follows from the $\mu_B\to-\mu_B$
symmetry of the QCD partition function. Solving the two conditions for
$T$ and $\mu_B^2$ one finds the position of the critical point
$(T(m_s),\mu_B^2(m_s))$ in terms of functions analytic in $m_s$.

de Forcrand and Philipsen determine the function $\mu_B^2(m_s)$,
or rather its inverse $m_s(\mu_B^2)$, for $\mu_B^2<0$ and then
analytically continue to real $\mu_B$. This way one could estimate
the position of the critical point in the $T\mu_B$ plane.

It is puzzling that the slope of the function $m_s(\mu_B^2)$ measured
in this way \cite{Philipsen,deforcrand} appears negligible in lattice
units and has a wrong sign\footnote{I.e., opposite to the sign implied
  by Fig.~\ref{fig:std-scenario}.}  after a translation to physical
units is applied. This leads the authors of
Refs.~\cite{Philipsen,deforcrand} to suggest an unusual scenario: a
new critical point is emerging on the phase diagram as the $(m_s,m_{ud})$
point is taken into the 1st order domain on the Columbia plot,
dragging a new line of 1st order transitions into the $T\mu_B$ plane.
An unusual feature of such a point worth pointing out is 
the positioning of the 1st
order line on the {\em high\/} temperature side of the critical point.
As emphasized in Ref.\cite{deforcrand}, these results should still be
subject to large uncertainties due to discretization and/or
finite volume errors and more refined simulations are needed before
physical conclusions can be drawn.

\subsection{Taylor expansion}

Taylor expansion in $\mu_B$ is another method to circumvent the sign problem.
Derivatives of pressure (or other thermodynamic quantities)
 are calculated at $\mu_B=0$ and
assembled into a Taylor series expansion to obtain dependence of that 
quantity on $\mu_B$ \cite{Ejiri,Gavai,Ejiri:2006ft}

\begin{figure}
  \centering

\includegraphics[width=0.4\textwidth]{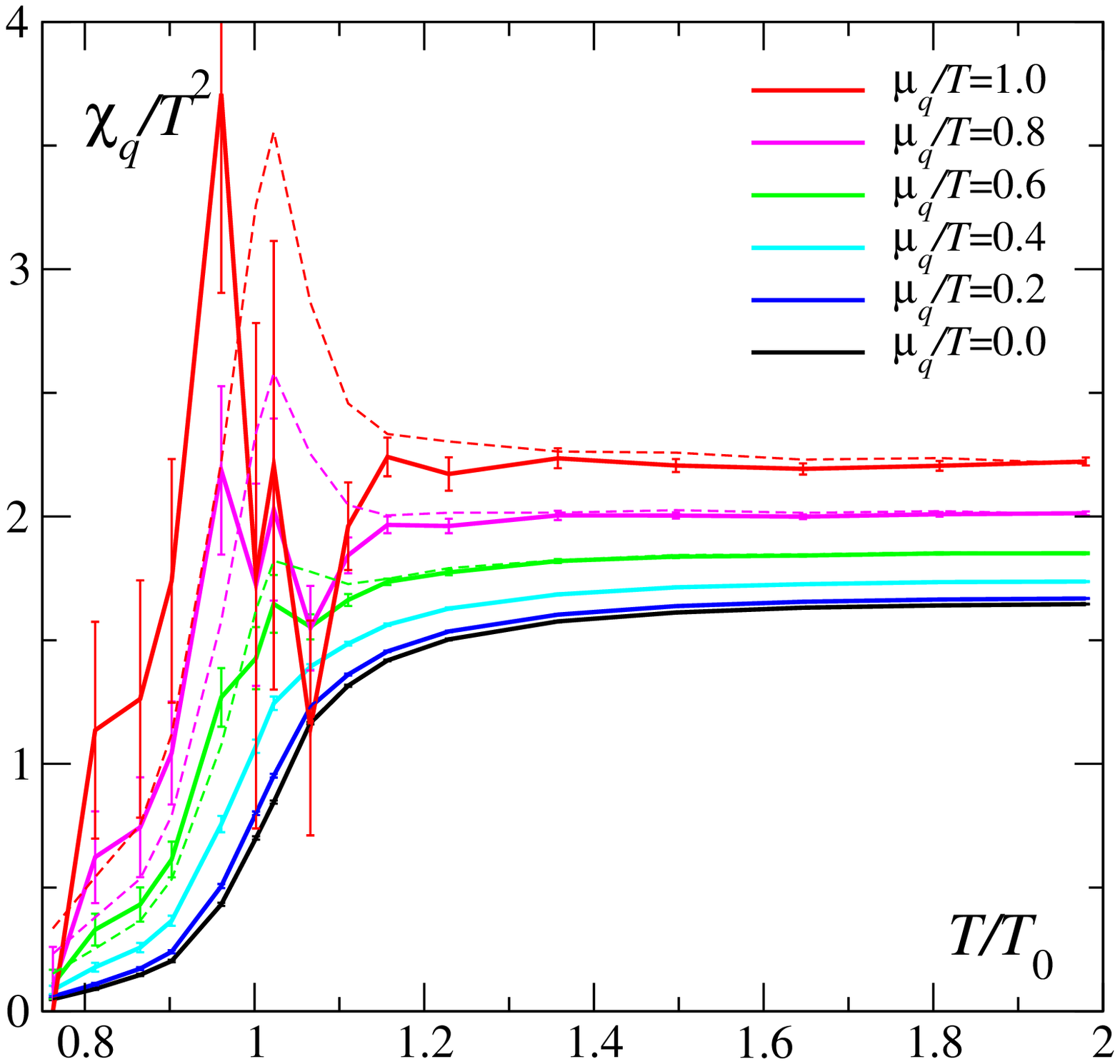}
\includegraphics[width=0.4\textwidth]{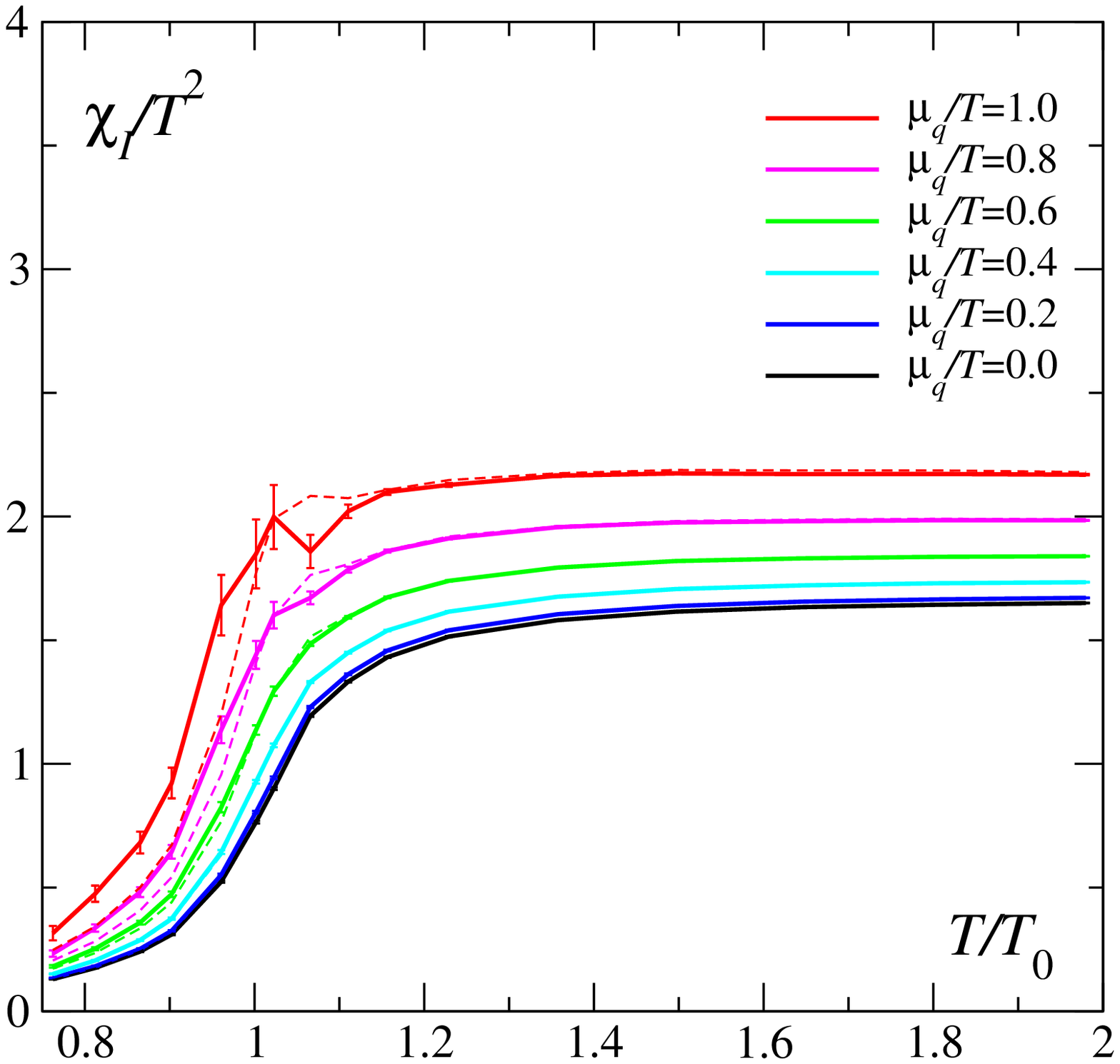}

\caption{Allton, {\em et al} \cite{Allton:2005gk}: 
peak in baryon number susceptibility $\chi_B=9\chi_q$ (left), but not in
  isospin susceptibility $\chi_I$ (right). See Ref.\cite{schmidt}
for updated figure.}
  \label{fig:allton-peak}
\end{figure}

Consistent with the existence of the critical point at finite $\mu_B$,
there is a noticeable rise in the baryon number susceptibility
$\chi_B$ -- see the peak on Fig.~\ref{fig:allton-peak}.
Such a peak should be expected
since the baryon number susceptibility diverges at the critical point.
On the other hand, the isospin susceptibility should not diverge
at the critical point, because the critical mode, $\sigma\sim
\bar q q$ is an isoscalar and cannot be excited by the operator of
isospin $I_3$ (while it is excitable by the operator of the baryon number).

The authors of Ref.~\cite{Allton:2005gk} 
caution against attributing the peak to the
critical point. Their reservation is due to the fact that 
the low-$T$ side of the peak is well described by a hadronic 
resonance gas model.
Nevertheless, the agreement with the resonance
gas does not necessarily mean that the rise of susceptibility cannot
be due to the critical point of QCD. On the contrary, viewing
resonance gas conceptually as a complementary (dual) description of
QCD one must conclude that resonance description must reproduce 
the same thermodynamic functions as
fundamental QCD description -- including the critical point.
Although the simple resonance gas model used in Ref.\cite{Allton:2005gk}
does not describe the critical point itself, it still might be describing the
onset of the critical behavior, just before the model breaks down.%
\footnote{
  A calculation illustrating this point has been reported earlier in
  Ref.\cite{Antoniou}: Improving the resonance gas description by a
  certain bootstrap procedure one obtains an equation of state which
  does have a critical point, similar to the van der Waals equation of
  state.}

Furthermore, the resonance gas model of Ref.\cite{Allton:2005gk} does not
describe the higher $T$ side of the peak. The model must break down as
the peak is approached from below, and is certainly not valid above
the peak, where a different description must be used. At the same time,
both sides of the peak receive a natural interpretation in terms of
the proximity of the critical point.

\subsection{Radius of convergence of the Taylor expansion}

At a fixed temperature, the convergence radius of the
Taylor expansion in $\mu_B$ is limited by the nearest singularity
in the complex plane of $\mu_B$. Assuming that at the temperature
$T_E$, at which the critical point $(T_E,\mu_E)$ occurs on the phase diagram,
this critical point is the nearest singularity to $\mu_B=0$, 
one could use Taylor expansion to determine $\mu_E$
\cite{Ejiri,Gavai,Ejiri:2006ft,schmidt}, if $T_E$ is known.

\begin{figure}
  \centering
  \includegraphics[width=.4\textwidth]{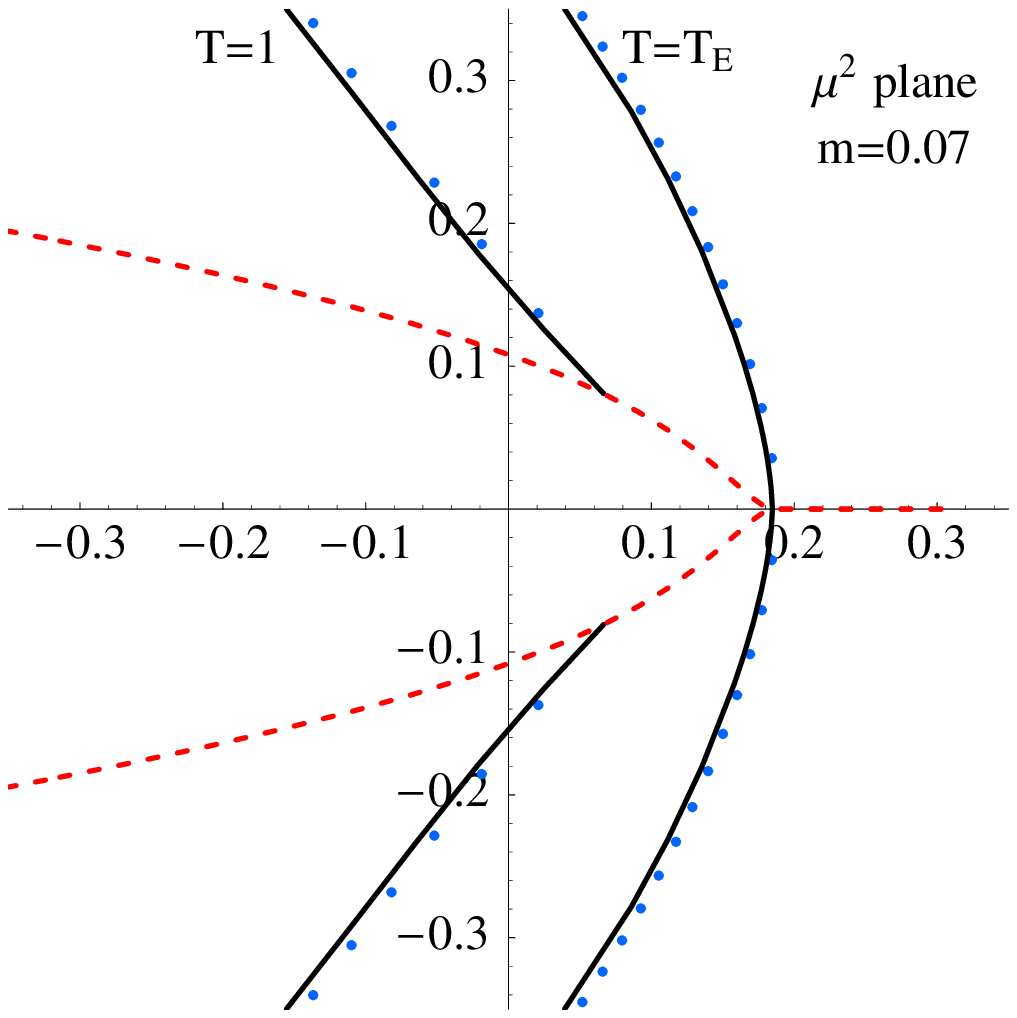}
  \hspace{.2\textwidth}
  \includegraphics[width=.3\textwidth]{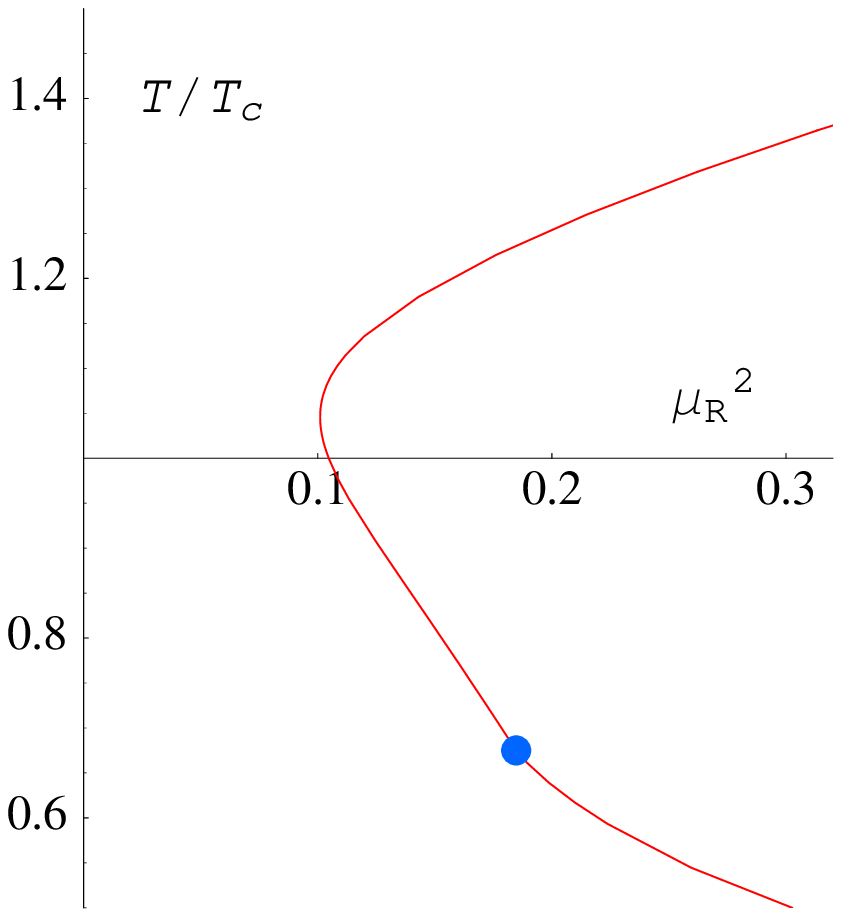}
  \caption{Ref\cite{Stephanov:2006dn}:
    The location of complex plane singularities (cuts in the
    thermodynamic limit)) in the random matrix model at two
    representative values of $T$: $T_c$ and $T_E$. At $T_E$ the
    branching points pinch the real axis -- this is where the critical
    point appears on the phase diagram. The trajectories of the
    branching points are shown by a dashed (red) curve. On the right
    the radius of convergence of the Taylor series, set by the
    distance from $\mu_B=0$ to the branching point, is plotted as a
    function of $T$ (the latter is along the ordinate to facilitate
    the comparison with QCD phase diagram). The critical point is shown.}
  \label{fig:convergence-radius}
\end{figure}

Assuming that the radius of convergence $\mu_R$ can be
approximated using the first few terms of the Taylor expansion, one can
plot $\mu_R$ as a function of $T$.  The main
remaining problem is to determine the value of $T_E$ i.e., to identify
at which value of $T$ the complex singularity reaches the real axis in
the $\mu_B$ plane. This question has been addressed using universality
arguments, as well as an example random matrix calculation in
Ref\cite{Stephanov:2006dn}.  The trajectory of the complex singularities
is illustrated in Fig.~\ref{fig:convergence-radius}.

Two conclusions can be made:
{\it (i)} The minimum value of the radius of convergence is {\em not}
achieved at $T=T_E$, but rather at a temperature close to the
temperature $T_c$
of the chiral transition at $\mu_B=0$.\footnote{In the chiral limit,
  the smallest value of $\mu_R$ is zero, and is achieved exactly
  at $T_c$ for $\mu_B=0$.
}
{\it (ii)} At $T_E$ the function $\mu_R(T)$ has a high order singularity.

It is unlikely that such a weak singularity alone can be used to
identify the value of $T_E$. This suggests that one should attempt to
extract more information from the Taylor series, for example, using
the complex phase of the $\mu_B$-plane singularity at given $T$. The
critical point could then be located by the condition that this
singularity is on the real axis. Such analysis would require observing
sign oscillations of the Taylor coefficients, and will require the
knowledge of the coefficients up to an order higher than available to
date.

\section{Scanning QCD phase diagram in heavy ion collisions}
\label{sec:scanning-qcd-phase}

Even though the exact location of the critical point is not
known to us yet, the available theoretical estimates
suggest that the point is within the region of the phase diagram
probed by the heavy-ion collision experiments. This raises the
possibility to discover this point in such experiments \cite{SRS}.

It is known empirically
that with increasing collision energy, $\sqrt s$, the resulting
fireballs tend to freeze out at decreasing values of the
chemical potential. This is easy to understand, since the
amount of generated entropy (heat) grows with $\sqrt s$
while the net baryon number is limited by that number in the initial nuclei.

The information about the location of the freezeout point for given
experimental conditions
 is obtained by measuring the ratios of particle yields
(e.g., baryons or antibaryons to pions), and fitting to
a statistical model with $T$ and $\mu_B$ as parameters.
Such fits are amazingly good \cite{BMRS}, and
the resulting points for different experiments are shown in
Fig.~\ref{fig:tmudat}.

As with any critical point, measurement of fluctuations can be used 
to determine when the system is in the vicinity of the critical point.
By measuring variables sensitive to the proximity of the critical
point as a function of monotonically increasing $\sqrt s$ of the
collision, and observing non-monotonic dependence, one discovers the
critical point \cite{SRS}. The values of $T\mu_B$ corresponding
to the freezeout at such a value of $\sqrt s$ give the coordinates
of the critical point.

As a concluding remark, it should be pointed out that the physics of
the critical point is universal (as far as slow and long distance
phenomena are concerned), which allows to define certain experimental
signatures independently of microscopic description.  However, the
position of the critical point on the phase diagram is determined by
microscopic physics and is not universal at all. This obviously makes
it very difficult to predict the coordinates of the critical point
reliably as it is evident in the scatter of predictions in
Fig.~\ref{fig:tmudat}.  On the other hand, the same fact should turn
the knowledge of the position of the critical point obtained on the
lattice, or in the experiment, into a powerful constraint on possible
models of QCD thermodynamics.

\acknowledgments
 This work is supported by the DOE (DE-FG0201ER41195), and
  by the A.P.~Sloan Foundation.

\end{document}